\documentclass[%
 reprint,
 amsmath,amssymb,
 aps,
floatfix,
]{revtex4-1}

\usepackage{graphicx}
\usepackage{dcolumn}
\usepackage{bm}
\usepackage[english]{babel}

\usepackage[utf8x]{inputenc}

\usepackage{geometry}		
\usepackage{graphicx}
\usepackage[above,below]{placeins}	
\usepackage{times}
\usepackage{float}
\usepackage{natbib}
\usepackage[bottom]{footmisc}

\begin{document}

\title{Student networks on online teaching due to COVID-19: Academic effects of strong friendship ties and perceived academic prestige in physics and mathematics courses}

\author{Javier Pulgar*}
\affiliation{Departamento de F\'isica, Universidad del B\'io B\'io, Concepci\'on, 4051381, Chile.}
\affiliation{Corresponding Author: jpulgar@ubiobio.cl}

\author{Diego Ram\'irez}
\affiliation{Magister en Ense\~nanza de las Ciencias, Universidad del Bío Bío, Chillán, 3800708, Chile.}

\author{Abigail Umanzor}
\affiliation{Magister en Ense\~nanza de las Ciencias, Universidad del Bío Bío, Chillán, 3800708, Chile.}

\author{Cristian Candia}
\affiliation{Data Science Institute, Facultad de Ingeniería, Universidad del Desarrollo, Las Condes, 7610658, Chile.}
\affiliation{Northwestern Institute on Complex Systems (NICO), Northwestern University, Evanston, IL 60208.}

\author{Iv\'an S\'anchez}
\affiliation{Departamento de F\'isica, Universidad del B\'io B\'io, Concepci\'on, 4051381, Chile.}


\date{\today}


\begin{abstract}
Collaboration among students is fundamental for knowledge building and competency development. Nonetheless, the effectiveness of student collaboration depends on the extent that these interactions take place under conditions that favor commitment, trust, and decision-making among those who interact. The sanitary situation and the transition to remote teaching has added new challenges for collaboration given that students' interactions are mediated by Information and Communication Technologies (ICTs). In this study we explore the effectiveness of different collaborative relationships on physics and mathematics, from a sample of secondary students from two schools located in rural and urban areas in southern Chile. We used Social Network Analysis (SNA) to map students' friendships relations, academic prestige, and collaboration on both courses. Later we combined the collaboration network with friendship and academic prestige on the course, to separate strong from weak friendship working ties, and those among students who enjoy or not academic prestige. Multiple linear regression models showed, on average, positive effects of collaboration on grades. Yet, when isolating the effects of the types of collaboration, the positive effects are observed only between those who display more strong friendship ties. Also, we found differences on the social networks and their effects over grades between both courses, presumably due to their pedagogical nature. With these results we contribute to the literature of collaboration and its effectiveness based on the nature of students' relationships, and advocate for the importance of instructional design in fostering appropriate motivations and guidelines for constructive collaboration in the classroom.
\end{abstract}

\maketitle

\section{Introduction}
Student collaboration is increasingly gaining attention for its benefits on learning and overall human development \cite{echeita2012, Barkley_ColTech, cerda2019}. Education scholars and policy makers have highlighted teamwork and social competencies as key abilities for life and work in the XXI Century \cite{Bao2019, pllegrino}. The emphasis on social skills is necessary to equip individuals with creative capacities in the face of complex and multidisciplinary challenges \cite{sawyer_edu_innovation}. The worldwide pandemic due to COVID-19, has without a doubt added external difficulties to the development of collaborative skills, as schools and universities were forced to move from face-to-face to remote instruction, and with the subsequent changes in students' roles and means for socialization and interaction. 

Remote teaching implies that students' communication is mediated via Information Communication Technologies (ICTs) \cite{Vonderwell2005,Traxler2018, panigrahi}, such as forums, chats, emails, or video conferences among other alternatives. Even though technologies are ubiquitous in today's society, the effectiveness of remote education strongly depends on their accessibility to technology, and on their readiness to navigate and deal with these with autonomy and self-regulation \cite{Hung2010, Kebritchi2017}. These conditions clearly affect how students' engage in forums or chats, instances that become proxies of social engagement and collaboration \cite{Romiszowski2004, Wise2013, Traxler2018}. Yet, engaging on effective collaboration implies finding the right partners to work with, which depends on assessing a myriad of variables related to the nature of the learning activity and its requirements \cite{pulgar_2021}, and the trust and commitment embedded in one's working relationships (i.e., friendship) \cite{Pulgar_TSC}. Here, the use of Social Network Analysis (SNA) affords relevant methodological and theoretical tools to capture the different types of social relationships they engaged on in the classroom, and how these foster (or hinder) their learning potential \cite{Brewe_Bruun16}.  

Consequently, and with the goal of further understand effective collaboration in physics and mathematics in remote classrooms, we turned to SNA \cite{Borgatti2013, carolan} to explore the academic effects of having working social ties with friends and/or other prestigious students. Based on students' social network we characterized different collaborative relationships depending on whether these occurred among friends (i.e., strong), in the absence of friendship (i.e., weak), and between proficient peers. This work contributes first, with a novel method to categorize students' working relationships following SNA, and second, with empirical evidence that shows the academic gains of various collaborative ties under distinctive teaching methodologies enacted in physics and mathematics (i.e., lecture-based and active learning). 

\section{Student Networks and Teaching Strategies}
Using SNA on education has enabled researchers to understand the benefits of student collaboration from the logic of social capital, in the sense that a greater number of social interactions -high network centrality- allows access various forms of resources and information, and the potential adoption of behaviours related to success \cite{Grunspan, Putnik}. Good performance has been found linked to students' social networks on face-to-face \cite{Bruun2013, Pulgar19, Putnik, Brewe_Sawtelle2012} and remote course \cite{Morris2005, Traxler2018,Dawson2008}, while covering a wide range of topics and strategies. Such findings align with the sociocultural view of cognitive and human development, as knowledge is constructed while learners work alongside others who could potentially expand the frontiers of their individual achievements, on what has been defined as the Zone of Proximal Development (ZDP) \cite{vyg}. 

The benefits of having multiple interactions with peers on the classroom, and becoming a central member of the network extends to other variables related to achievement, such as retention \cite{Williams,Zwolak_Zwolak, Zwolak}, and self-efficacy, or the believe on one's abilities to successfully meet the academic and performance expectations \cite{Dou2016}. Besides, programs grounded on the principles of collaborative learning have shown important results for developing social skills \cite{Carrasco2018}, and trust among team members, which increases the likelihood of group effectiveness \cite{leon2017}. 

Nonetheless, more recent evidence on SNA on university physics courses suggests that the academic gains from having multiple working ties depend on the nature of the learning task \cite{Pulgar19}. Accordingly, a great number of social ties relate to having a worse performance on well-defined physics problems (e.g., textbook problems), whereas open-ended and creative activities benefit from multiple social ties connecting different groups (i.e., brokering knowledge \cite{Burt2004,Pulgar19}). Additional results on primary education found that having multiple ties is again linked with poor performance, while the reciprocity of students' relationships becomes the necessary condition for higher achievement \cite{Candia}. 

Furthermore, according to evidence on sociology and networks, the nature of the social relationship plays a key role in acquiring and developing knowledge depending on its complexity \cite{Granovetter, hansen}. For instance, new tacit, factual or \textit{knows what} type of information is easily accessible through weak social ties, that is, among individuals who do not share a sense of commitment nor are deeply embedded into the relationship \cite{Granovetter}. The key consideration is that weak ties are observed between individuals who do not belong to one's cohesive groups -governed by strong relationships-, and are therefore preferable for accessing simple new ideas outside such groups. On the contrary, strong ties are better suited for learning complex and non-tacit ideas, because developing such knowledge requires a social commitment and a common language to transfer the intricacies of this new information \cite{hansen}. Such learning considerations are frequently observed among individuals who share a strong relationship, like friends. 

All of the above stresses out the challenges of guiding students' collaboration in the classroom, taking into account the nature of the tasks and information to be learned. These complexities are now relevant in this transition to remote teaching, where students' interactions are mediated by ICTs \cite{Traxler2018}. Studies on remote learning have conceptualized students' course participation using traditional methods, such as the number of posts on online forums \cite{Romiszowski2004, Vonderwell2005}, the time dedicated to reading comments \cite{Hrastinski2009,Wise2013}, or via content analysis of the posts written \cite{deweber}. Recently, researchers have used SNA to explore different forms of participation on online courses. For instance, Traxler, Gavrin and Linden \cite{Traxler2018} conceptualized students' co-occurrence of posts on discussion forums, which were defined as the links on the participation network, and later used to determine and understand students' centrality based on their access to the ideas presented. Under this condition, higher access to information on forums, that is, high network centrality is positively correlated with academic success \cite{Traxler2018}. Through similar network methods, an additional study has found positive correlations between students' participation on online courses -centrality- with their sense of belonging \cite{Dawson2008}. Yet, the centrality on this participation network offers positive effects specially to high performance students, as content analysis has shown that their discussions circled around conceptual aspects of the course content, whereas the discussion networks formed by low performing students tend to focus on practical and superficial elements of curriculum \cite{dawson2010}.

From the perspective of learning methodologies, active classrooms have shown higher levels of social interactions compared to traditional lecture-based instruction \cite{traxler2020,Commeford2020,Pulgar19, Pulgar_Sochifi,Brewe_Kramer2012}. Student-centered classrooms favor autonomy over the learning process through peer interaction, because these tend to include activities designed to encourage decision-making, content manipulation and communication, thus allowing students to perceive learning as a construction mediated by collaboration \cite{pulgar_2021, Pulgar19}. The social and cognitive attributes of activities implemented on active learning methodologies are also recommended on virtual classrooms \cite{chametzky, Luyt2013, Niess2013, gupta2013}, because these encourage higher levels of academic achievement and digital competencies \cite{merono2021}. 

Finally, the work of Johnson, Johnson and Holubec\cite{Johnson_Johnson} on student collaboration adds fundamental conditions for group effectiveness: positive interdependence, or the belief that the success is a collective rather than individual effort; and personal responsibility for learning and engaging on one's tasks. According to social network research on education, the above conditions are modulated by the nature of the learning tasks. For instance, close-ended well-structured physics problems hinders positive independence, as students report addressing these activities by engaging in social interactions aimed at finding the right equation or variable to use \cite{pulgar_2021}. On the contrary, real-world problems \cite{Fortus} such as generative activities \cite{pulgar_2021}, motivate collaborative strategies that align with the mentioned characteristics for effectiveness. Among the reasons why one would witness such a different form of collaboration on these two sets of problems -close-ended and generative- links with their embedded attributes. Accordingly, well-structured and close-ended problems are perceived as non-additive tasks \cite{Steiner1966}, given that when worked on groups these tend to be solved by the most capable team member. These problems also relate to rather simplistic cognitive processes (e.g., apply content) according to a taxonomy of physics problems developed by  Teodorescu and colleagues \cite{teo}. Differently, generative activities are intrinsically additive \cite{Steiner1966}, because they require collective efforts for content manipulation, decision-making and manipulation, and thus demand high-level cognitive skills \cite{Anderson2001, teo}.
 
\section{Methods}

\subsection{Research Context}
The study was conducted during the second semester of 2020 on a sample of secondary students from parallel classes (A \& B), and from two schools in southern Chile (Sch-1 \& Sch-2). This exploratory research focused on the courses of physics and mathematics. Table \ref{Tabla1} summarizes the schools' characteristics, such as the student population, research participants, location and education. Research participants gave consent via an online survey designed for this study, and where we explain the analytical procedures along with privacy considerations. A total of 101 students agreed to be involved in the study, with 45\% of female participants.

\begin{table}
\begin{center}
  \caption{School attributes and characteristics \label{Tabla1}}
    \centering
 \begin{tabular}{p{0.9cm}cc p{1.05cm} p{3.15cm}}
\hline\hline
School       & Students & Participants & Location & Education\\
\hline 
Sch1 & \centering 650   & \centering 55 & Rural    & Scientific, Humanistic \& Technical\\
Sch2 & \centering 906   & \centering 46 & Urban & Scientific \& Humanistic \\
\hline
\end{tabular}
\end{center}
\end{table}

During 2020 and due to the COVID-19 pandemic, both schools adapted their teaching methodologies by introducing different ICTs. Initially, on School 1 teachers began the academic year with asynchronous sessions using content videos shared via text messaging apps (e.g., WhatsApp). After two months, School 1 implemented synchronous 40 min sessions organized on MEET and CLASSROOM learning management systems (LMS). This scenario differed from School 2, where remote teaching began with 1-2 hours synchronous course sessions a week after the school canceled face-to-face instruction. These sessions, similar to School 1 were organized on the same LMS. 

In addition to the initial organizational differences, both schools organized and taught physics and mathematics content following opposite learning methodologies. In School 1, lecture-based instruction was the main methodology on both courses, followed by problem solving sessions where students faced well-structured activities (e.g., textbook problems), and were later assessed on their individual work. In School 2, the mathematics course met the characteristics of School 1 with 3 weekly classes, but with the difference that sessions were 1-2 hours long compared to 40 min on School 1. Finally, physics on School 2 was based on active learning methodologies, including generative activities \cite{pulgar_2021}, where students, for instance, working on groups designed videos centered around physics concepts and phenomena. The physics classes met once a week for 40 min and 1 hour respectively for Schools 1 and 2.

\subsection{Data Collection}
Because the study aims to determine the academic effects of different collaborative relationships on school physics and mathematics, we collected final grades from the year prior to the study (2019), along with students' grades on the second semester of 2020. In 2020, students' performance on both courses on School 1 accounted for in-class activities (e.g., problem solving) and individual testing. This is similar to how mathematics' grades were obtained in School 2. Lastly, physics grades on School 2 consisted on the assessment of the generative activities (e.g., content videos and physics posters), which consisted on grading their structure and organization, language and theory, including a peer evaluation of the groups' performance. 

Finally, student collaboration was measured through online network surveys, shown to be a reliable mechanism for network mapping \cite{Borgatti2013}. This survey was designed to gather different social relationships: friendship ties; prestige on physics and mathematics; and collaboration on physics and mathematics. The survey was constructed using the class' roster, which has been utilized in previous studies to facilitate responses for network measuring \cite{Brewe_Bruun16, traxler2020, Pulgar19}.

\subsection{Data Analysis}
From the network survey we identified key variables to characterize the groups in terms of their friendship relations, prestige and collaboration on both physics and mathematics courses. The survey yields an indirect network, that is, where ties are not necessarily reciprocal (e.g., A selects B as a friend, but B does not select A as a friend). Tables \ref{Tabla2} summarize the mean of key descriptive network variables for each of the social dimensions measured on the survey. The variables include the number of students per group, or nodes, the total number of ties observed, the average number of ties or average degree, and network density --percentage of observed ties considering 100\% as if all nodes on the network were connected to each other \cite{Borgatti2013,carolan}.

\begin{table*}[t!]
\begin{center}
  \caption{Descriptive statistics of directed networks: Friendship, Physics Prestige, Collaboration on Physics, Mathematics Prestige, and Collaboration on Mathematics.\label{Tabla2}}
  \centering
 \begin{tabular}{lcccccccccccccccc}
\hline\hline
           &          & \multicolumn{3}{c}{Friendship}   & \multicolumn{3}{c}{Physics Prestige} & \multicolumn{3}{c}{Collaboration on Physics} & \multicolumn{3}{c}{Math Prestige} & \multicolumn{3}{l}{Collaboration on Math} \\
Group     & Nodes & Ties & Degree$^*$ & Density        & Ties & Degree$^*$ & Density              & Ties & Degree$^*$ & Density & Ties & Degree$^*$ & Density & Ties & Degree$^*$ & Density  \\
\hline 
Sch1-A &  29   &  132 & 4.55 & 0.16 & 54&  1.86& 0.07&  43& 1.48&  0.05 & 84   &  2.9 & 0.1    & 34 & 1.17&  0.04  \\
Sch1-B &  26   &  95 &  3.65 & 0.15 & 67&  2.58& 0.1&  49& 1.89&  0.08  & 88     &  3.39 &  0.14 & 37 & 1.42&  0.06 \\
Sch2-A &  23   &  194 & 8.44&  0.38 & 126& 5.48 & 0.25&  52& 2.26& 0.1 & 138    &  6 & 0.27    & 55 & 2.39& 0.11 \\ 
Sch2-B &  23   &  151 & 6.57  & 0.3 & 128&  5.67& 0.25&  55&  2.39&  0.1 & 157     & 6.38 & 0.31  & 61 & 2.65& 0.12 \\
\hline
\multicolumn{17}{l}{$^*$: Average degree, or average number of incoming plus outcoming ties per student on the class.} 
\end{tabular}%
\end{center}
\end{table*}

\begin{figure*}[!htbp]
\centering
  \includegraphics[width=0.5\textwidth]{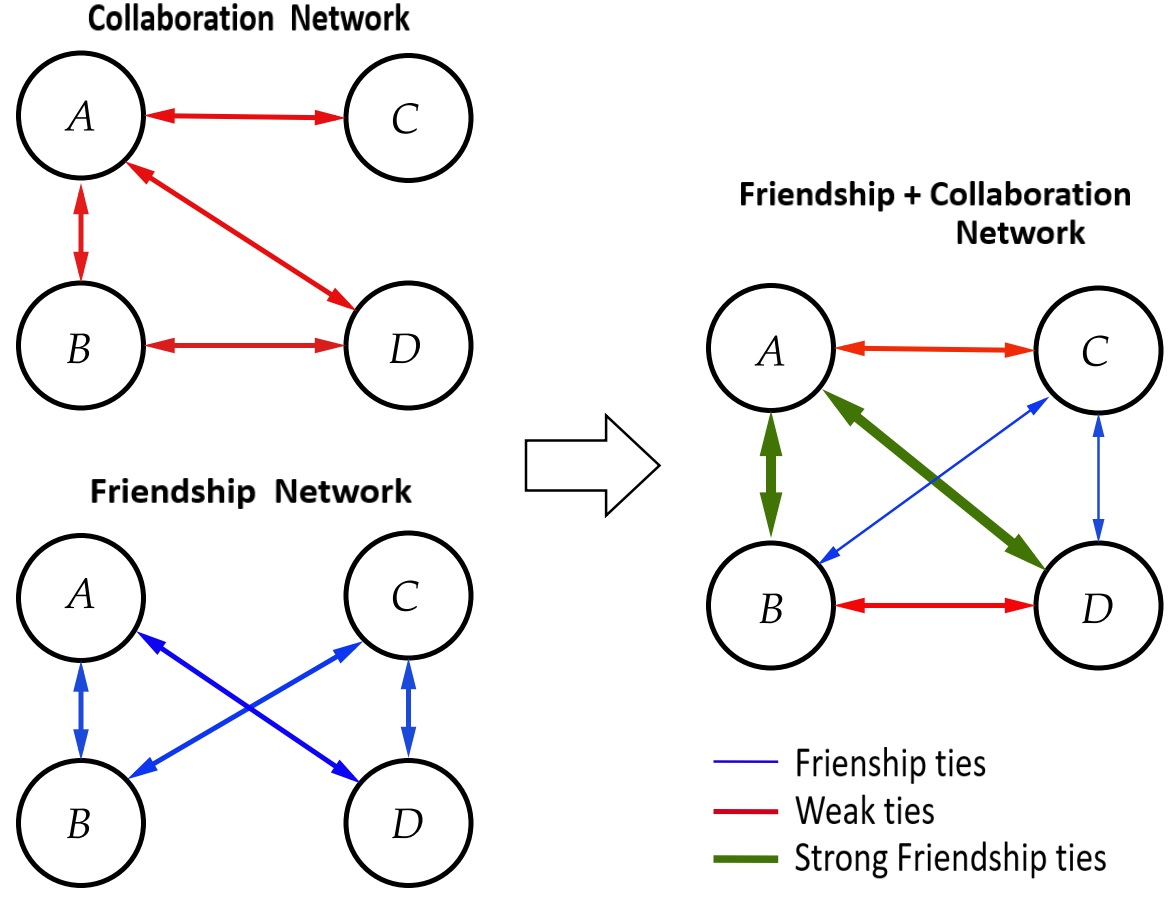}
  \caption{Diagram shows the methodology utilized to identified and construct collaborative variables number 2, 3, 4 and 5 described in Table \ref{Tabla3}. \label{methodology}}
\end{figure*}

We operationalized collaboration as degree centrality on the collaboration network, or as the total number of ties. Degree centrality accounts for both incoming and outgoing ties, named indegree and outdegree centrality respectively. The former indicates the number of ties headed from nodes on the network towards a focal actor, whereas the latter accounts for relationships declared by the focal actor towards other nodes on the network. Besides this definition of the variable \textit{Collaboration} (see Table \ref{Tabla3}), we combined friendship, prestige and collaboration networks to identify different types of collaborative relationships among students. We turned to the strength of ties terminology used on the social network literature to differentiate between strong and weak collaborative ties \cite{Granovetter, hansen}, based on whether the observed relationship occurs among students who are or not friends. The process followed to construct these variables is depicted on Figure \ref{methodology}, and shows how we combined two networks (e.g., collaboration and friendship) to define a third network of interactions but with weighted ties, given the added friendship attribute. In the diagram these weights represent either collaboration between friends (i.e., strong ties), or between students no are not friends (i.e., weak ties), or simple friendship ties. Because our interest is placed on different types of collaboration, we counted the number -degree centrality- of these collaborative ties and saved them as individual attributes (e.g., node A has 4 strong and 2 weak ties, considering the incoming plus the outgoing ones). Table \ref{Tabla3} describes the five collaboration variables constructed and used for the analysis. 

\begin{table*}[t!]
  \caption{Types of collaborative relationships and definition \label{Tabla3}}
  \centering
 \begin{tabular}{p{4cm} p{12cm}}
\hline\hline
Type of Collaboration                           & Definition \\
\hline

1. Collaboration                       & Degree centrality of the collaboration network, or number of collaborative ties.\\

2. Strong	     & Number of collaborative ties between students who are friends. Friendship is observed on the network when at least one of the actors on the dyad indicates the other as a friend (e.g., A → B). \\

3. Weak     & Number of collaborative ties between students who do not declare themselves as friends.\\

4. Strong w/ Academic Prestige    & Number of collaborative ties between friends and who are perceived by the other as a good student on the course. Friendship is observed on the network when at least one of the actors on the dyad indicates the other as a friend (e.g., A → B). Similarly, prestige comes from its respective network and when at least one of the actors on the dyad selects the other as a good student on the course (ej., B → A, then B enjoys academic prestige).\\

5. Weak w/ Academic Prestige	& Number of collaborative ties between students who do not declare friendship ties among themselves, yet one of the actors on the dyad declares the other as a good student.  \\

\hline
\end{tabular} %
\end{table*}

It is worth mentioning that each of the five variables shown on Table \ref{Tabla3} were built for their respective course (i.e., physics or mathematics). For instance, the collaboration network on mathematics, along with academic prestige on the same course and friendship were combined to determine the five collaborative ties described above for mathematics, and the same occurs for physics with its respective networks. Later, we used ordinary least squares multiple regressions models (OLS) to predict grades on both courses, while using collaboration variables as the main predictors. In order to isolate the effect of collaboration over grades, we include control variables like grades on the year prior to the study, School, and gender (female). Finally, as a robust check, we fitted hierarchical linear models to account for the class and schools' variance. Results here did not differ from the multiple regression models. 

\section{Results}

\subsection{Collaboration and Grades on Physics}

\begin{figure*}[!htbp]
\centering
  \includegraphics[width=1\textwidth]{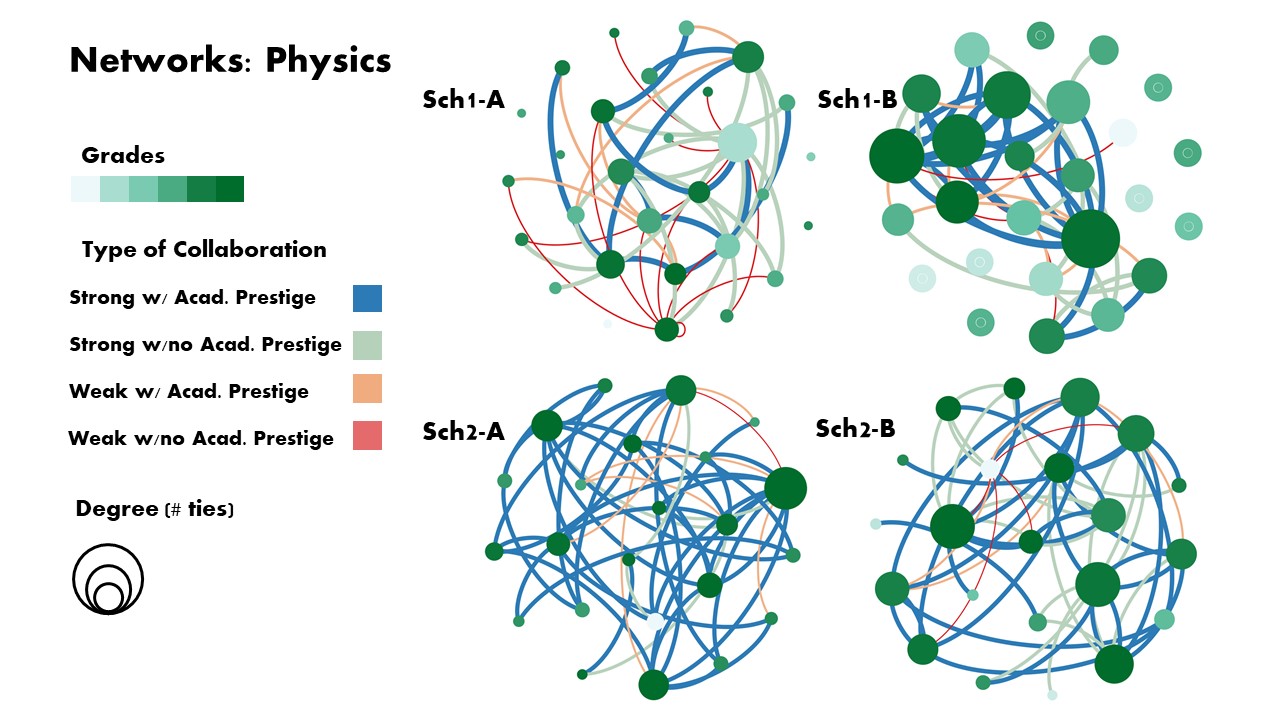}
  \caption{Collaboration network for physics. Node color represents academic performance, while its size is degree centrality --number of collaborative ties. Edge  or link colors depict the type of collaborative relationship.\label{physics_networks}}
\end{figure*}

Figure \ref{physics_networks} depicts the collaboration networks of all four classes from Schools 1 and 2. The networks show four types of ties based on whether students display strong or weak relations with others, who at the same time might or not enjoy academic prestige in physics. Beside, the network depiction illustrates grades in shades of green, as well as degree centrality as the nodes' size. 

A first look of these networks allows a visual representation of the information contained on Table \ref{Tabla2}, where classes Sch1-A and B have low network density, represented by the reduced number of ties compared to School 2. Plus, there are differences between groups in School 1, as in class A we observe an abundance of social relations of diverse nature spanning most of the students, whereas class B seems to be governed by strong ties with prestige (blue links) and with many isolated members. In this last group, connected students with high centrality appeared with the highest grades, differently from Sch1-A, where good grades are not an exclusive attribute of central individuals. Finally, both networks on School 2 show an even distribution of good grades, while their ties are mainly strong.

\begin{figure*}[!htbp]
\centering
  \includegraphics[width=0.8\textwidth]{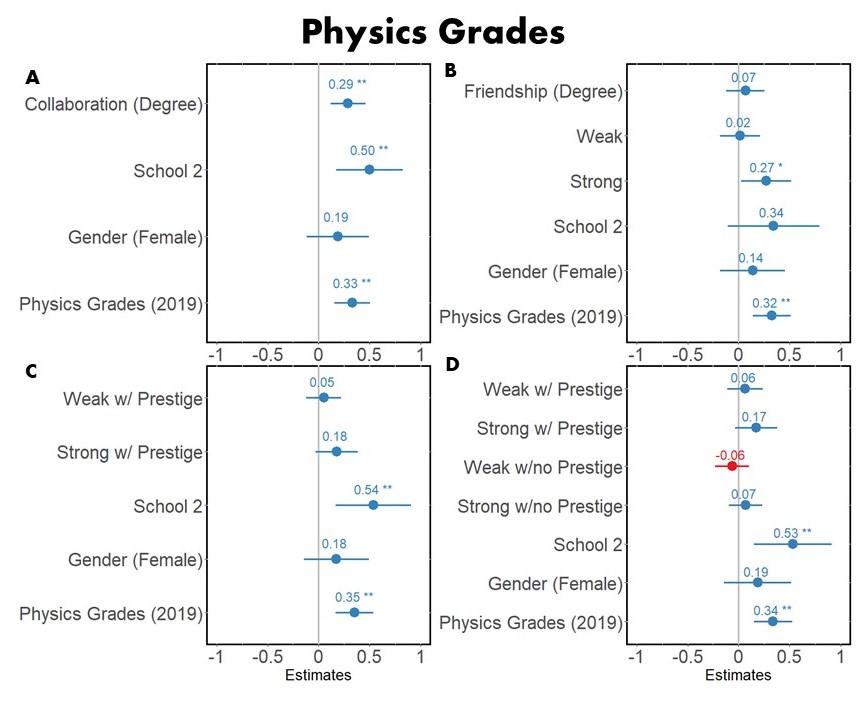}
  \caption{Multiple linear regression models for physics grades. Positive coefficients are depicted in blue, and negative coefficients in red. Note: * $p < .05$ y ** $p < .01$.\label{mod_phys}}
\end{figure*}

The multiple regression models fitted to predict physics grades are shown on Figure \ref{mod_phys}. On model A we used degree centrality of the collaboration network as the main predictor, yielding a positive coefficient (.29, $p < .01$) and thus, suggesting that more working partners foster success in physics. When differentiating the effects of types of collaboration, model B shows a null effect for weak ties, while having strong relationships indicates a positive predictive value over physics performance (.27, $p < .05$). When adding academic prestige, models C and D show no significant gains of working alongside others who are perceived as good physics students. Finally, it is worth highlighting the predictive value of prior grades on physics across all four models, and the mean difference between Schools' 2 and 1 in favor of the former.

\subsection{Collaboration and Grades on Mathematics}

\begin{figure*}[!htbp]
\centering
  \includegraphics[width=1\textwidth]{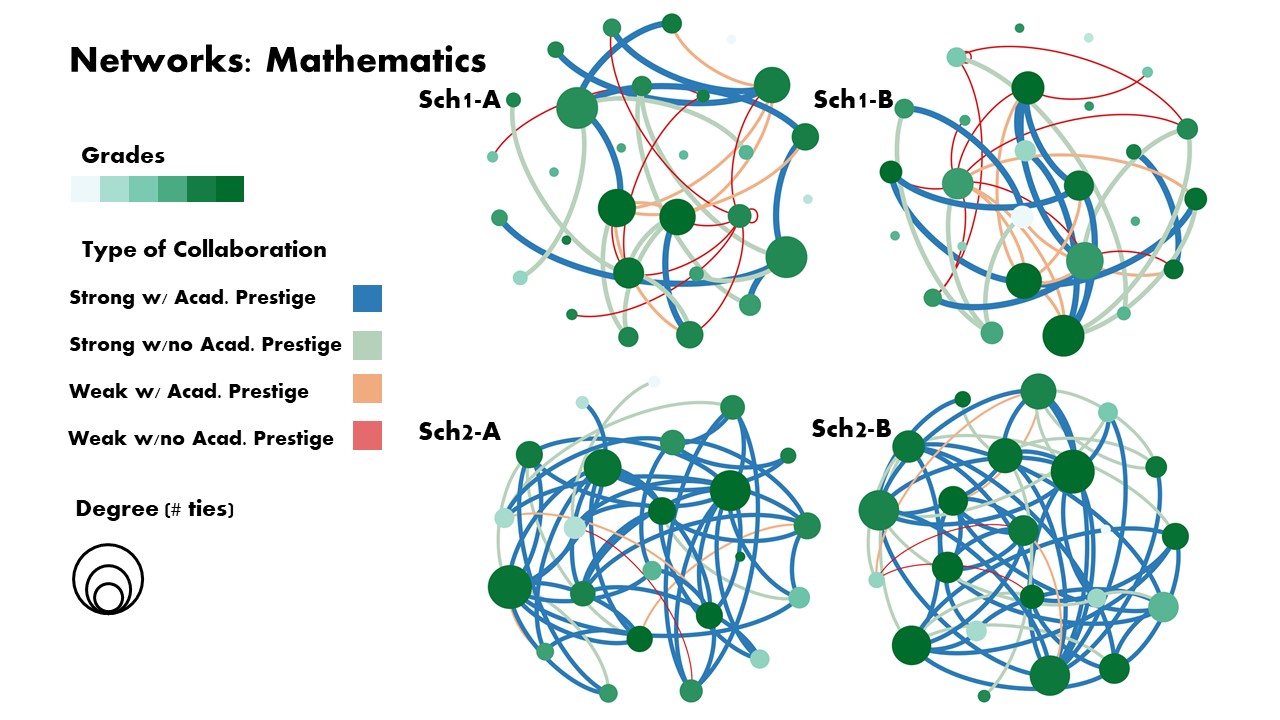}
  \caption{Collaboration network for mathematics. Node color represents academic performance, while its size is degree centrality --number of collaborative ties. Edge  or link colors depict the type of collaborative relationship.\label{math_networks}}
\end{figure*}

Similar to what was shown in physics, the mathematics networks on figure \ref{math_networks} depicted the different levels of social connectivity between groups and schools. However, across groups, nodes with higher degree centrality enjoy better grades. Additionally, in both classes from School 1 there are an important number of weak ties between students not recognized as prestigious, plus a similar number of isolated nodes. Importantly, good grades are observed among those who display strong collaborative ties, with either other good students or not. Differently, in School 2 it is possible to distinguish mainly strong social ties among students who get a higher grade, whereas low performance seems linked to weak interactions. 

\begin{figure*}[!htbp]
\centering
  \includegraphics[width=0.8\textwidth]{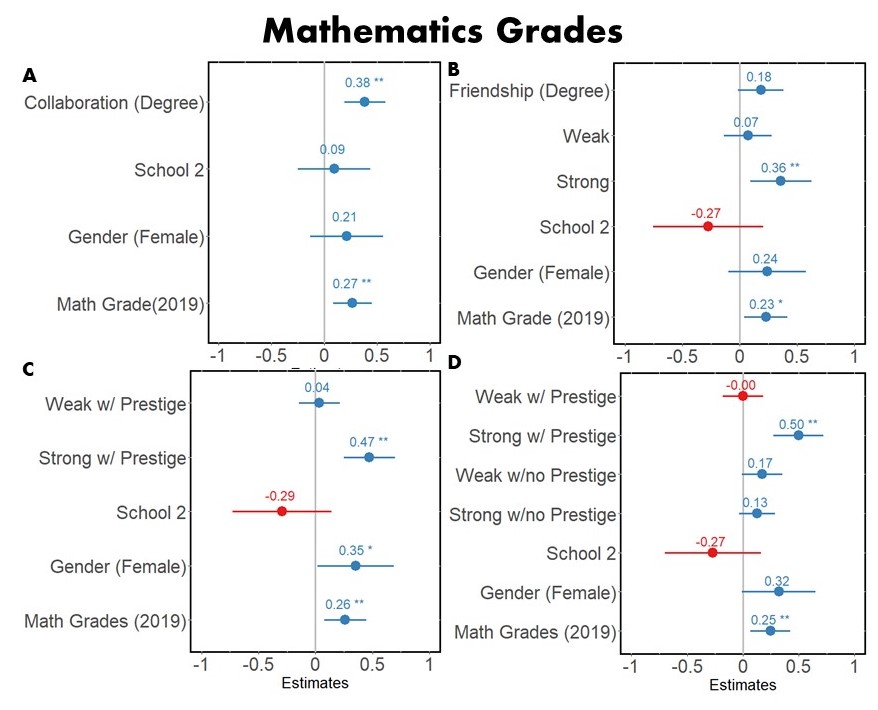}
  \caption{Multiple linear regression models for mathematics grades. Positive coefficients are depicted in blue, and negative coefficients in red. Note: * $p < .05$ y ** $p < .01$..
\label{mod_mat}}
\end{figure*}

Figure \ref{mod_mat} shows the multiple regression models for mathematics grades. Similar to physics, there is evidence on model A that having more collaborative partners (degree) relates to better outcomes on math (.38, $p < .01$). Yet, these academic gains are observed among those who interact with their friends (strong ties, .36, $p < .01$) rather than among students who work away from the close social circles (weak ties), as suggested by coefficients on model B. The benefits of strong ties strengthen when adding academic prestige into the dyadic relationship, with coefficients of .47 and .5 ($p < .01$) on models C and D respectively. Finally, female students score significantly better on mathematics according to the last two models, while again performance in the previous year resulted in a significant predictor of success.

\section{Discussion and Conclusions}
Results show clear differences in the density of collaboration and friendship ties between both schools. This contrasting scenario might be attributable to the socioeconomic and cultural characteristics found on both student populations. Even though we do not count with information about students' social or economic capital, schools' geographical locations rural and urban allow us to extrapolate certain comparative advantages relative to access to internet and communication technologies, which might have facilitated or hindered students' interactions for miscellaneous purposes. According to the literature on distance education and ICTs, success on remote teaching depends highly on students' accessibility and their readiness to use digital tools \cite{Hung2010, Kebritchi2017}. Consequently, one might think that students from School 1 (rural) experienced higher limitations on accessing internet and interacting with ICTs, yielding to less ICTs' mediated communication compared to School 2, and resulting in lower network density, even though the former counts with a larger number of students per class. 

Besides, the teaching methodologies enacted by teachers in each of these classrooms might also allow us to interpret the observed differences on the number of social relationships across schools. Mathematics at both institutions and physics at School 1 were taught in a traditional fashion, with lectures and well-defined problems, similar to textbook activities students worked on individually. As found in the literature, well-defined problems introduced as group activities tend to favor isolated work and diminish collaboration due to its lack of positive interdependence \cite{pulgar_2021, Steiner1966}. This pedagogical model centers around the teacher rather than their students, and encourages on them a passive role grounded on individual performance \cite{pulgar_2021}, and might have hampered their motivation and the need to expand their existing collaboration and friendship networks for success. Therefore, such learning  might help them to fortify rather than diversify students' social networks in the classroom. 

Conversely, the teaching methodology implemented on the physics course on School 2 was active and student-centered, and where participants engaged in collective creative processes for instance, to develop physics videos and posters. These educational conditions might have contributed to the density of collaborative interactions, given that the generative group activities enacted here are characterized by encouraging high levels of decision-making \cite{teo} and positive interdependence among team-members \cite{Johnson_Johnson}. Further, research evidence has shown that active learning methodologies are preferred over lecture-based classrooms at fostering social relationships \cite{Brewe_Kramer2012}. Finally, and because collaborative, friendship and prestige networks were measured months after the first generative activity was implemented, its social effects could have extended to the mathematics course, and the reason why both disciplines on School 2 display high network density.

The multiple regression models show the expected results of having multiple working peers in the classroom, which align with the benefits of social integration on education \cite{Tinto}. The observed learning gains are also consistent with sociocultural theory \cite{vyg} and the principles of social capital \cite{zhao2010}, in the sense that social relationships enable access to information and ease collective learning, on this context, pertaining to physics and mathematics knowledge. Nonetheless, when examining the academic effects of different collaborative relationships, strong friendship ties showed an advantage compared to weak ties observed among those who do not declare to be friends. This evidence suggests that the physics and mathematics information and behaviors needed for a successful performance is better developed by working with closely related individuals. Consequently, the content addressed and assessed on both courses presents enough complexity that its learning requires strong friendship ties, as suggested by  Hansen \cite{hansen}, and because individuals on cohesive networks based on trust and mutual commitment share common forms of communication and understanding to invest time and energy on their collective growth. These claims are also supported by the null effects of weak ties across all regression models, given that these social relationships are preferred for learning simple and tacit knowledge \cite{Granovetter}, a type of information that does not seem relevant in these learning contexts.     

The evidence related to academic prestige requires an important consideration. According to the regression models, collaboration with friends who are perceived as good students (strong with prestige) becomes a positive predictor on mathematics, yet its effect is null on physics. Such differences might be attributable to the teaching methodologies and assessments enacted on both courses. On the physics course on School 2, students worked on groups to address collaborative activities that have shown to encourage social interactions \cite{traxler2020, Commeford2020, Pulgar_Sochifi,Brewe_Kramer2012}, which could have enable students to develop friendship and perceptions of proficiency beyond their initial social networks. This teaching orientation has shown to hinder the advantages of academic hierarchies, as the collaboration networks are not governed by clusters of proficient friends. In contrast, and as mentioned earlier, the individualistic teaching methodology observed in mathematics seems to accentuate the existing networks and the advantages of already having working ties with good students, thus fortifying the social and academic hierarchies in the classroom.   

Consequently, due to the creative and collaborative nature of the activities designed for physics in School 2, it is possible that participants were able to extend their own conceptions of proficiency beyond the traditions of solving well-structured problems, thus recognizing a larger number of good students in such a category. Such re-conceptualization of academic prestige enabled them to limit the gains of academic hierarchies, in the sense that proficiency is no longer exclusively associated with having good grades. This argument has been explored in the physics education literature, where prestige is associated with activities centered around textbook-types of problems, yet not with creative and generative tasks \cite{Pulgar19}. 

Plus, results allow us to characterize the observed network differences in terms of variables like belonging, trust and self-efficacy and the role teaching methodologies have on fostering them. According to the evidence, active learning methodologies foster higher degree centrality than lecture-based instruction. Research studies have suggested that under the former learning conditions students achieve higher sense of belonging \cite{Dawson2008}, and self-confidence to face academic tasks \cite{zhao2010, Zwolak_Zwolak}. This interpretation adds up the findings and support for the design and use of students-centered and collaborative teaching strategies, for both remote \cite{chametzky,Luyt2013, Niess2013, gupta2013} and face-to-face classrooms \cite{pulgar_2021}.

Finally, the limitations of the study relate to the absence of information for instance, on internet connectivity, accessibility and teacher and student training on ICTs, given the major role these play in remote education. Besides, we recognize that the conducted study does not account for students' personal experience during the transition from face-to-face to online instruction. Here, qualitative information and analytical tools might have provided valuable information to either support or reject the interpretations described in the paragraphs above. Furthermore, the study pretends to encourage the research and teaching community to reflect and  pursue new studies and interventions on the interplay between students' networks, teaching methods and learning. This avenue of research is promising, particularly since social and collaborative skills are nowadays necessary for human development and to face the complex challenges of the XXI Century \cite{Bao2019, sawyer_edu_innovation}.

\section*{Acknowledgement(s)}

This material is part of the research project 2020217 IF/F: \textit{Knowledge and collaboration: a social network approximation to understand collaboration practices university physics classrooms}, funded by Vicerrecto\'{i}a de Investigaci\'{o}n y Postgrado, Universidad del B\'{i}o B\'{i}o, Concepci\'{o}n, Chile.

\pagebreak
\bibliographystyle{unsrt}
\bibliography{ref}

\begin{thebibliography}{10}

\bibitem{echeita2012}
G.~Echeita.
\newblock El aprendizaje cooperativo al servicio de una educación de calidad:
  Cooperar para aprender y aprender a cooperar.
\newblock In J.~C. Torrego and A.~Negro, editors, {\em Aprendizaje Cooperativo
  en las aulas: Fundamentos y Recursos para su Implantación}. Alianza
  Editorial, Madrid, España, 2012.

\bibitem{Barkley_ColTech}
E.~Barkley, C.~H. Major, and K.~P. Cross.
\newblock {\em Collaborative learning techniques: A handbook for college
  faculty}.
\newblock Jossey-Bass a Wiley Brand, San Francisco, CA, USA, 2014.

\bibitem{cerda2019}
G.~Cerda, C.~P\'{e}rez, P.~Elipe, J.A. Casas, and R.~Del~Rey.
\newblock Convivencia escolar y su relación con el rendimiento académico en
  alumnado de educación primaria.
\newblock {\em Revista de Psicodidáctica}, 24(1):46--52, 2019.

\bibitem{Bao2019}
L.~Bao and K.~Koenig.
\newblock Physics education research for 21st century learning.
\newblock {\em Disciplinary and Interdisciplinary Science Education Research},
  1(2):1--12, 2019.

\bibitem{pllegrino}
J.W. Pellegrino and M.L. Hilton.
\newblock {\em Education for Life and Work: developing transferable knowledge
  and skills for 21st century}.
\newblock The National Academies Press, New York, 2003.

\bibitem{sawyer_edu_innovation}
R.K. Sawyer.
\newblock Educating for innovation.
\newblock {\em Thinking Skils and Creativity}, 1(1):41--48, 2006.

\bibitem{Vonderwell2005}
S.~Vonderwell and S.~Zachariah.
\newblock Factors that influence participation in online learning.
\newblock {\em Journal of Research on Technology in Education}, 38:213--230,
  2005.

\bibitem{Traxler2018}
A.~Traxler, A.~Gavrin, and R.~Lindell.
\newblock Networks identify productive forum discussions.
\newblock {\em Physical Review Physics Education Research}, 14:020107, 2018.

\bibitem{panigrahi}
R.~Panigrahi, P.~R. Srivastava, and D.~Sharma.
\newblock Online learning: Adoption, continuance, and learning outcome—a
  review of literature.
\newblock {\em International Journal of Information Management}, 43:1--14,
  2018.

\bibitem{Hung2010}
M.~Hung, C.~Chou, C.~Chen, and Z.~Own.
\newblock Learner readiness for online learning: Scale development and student
  perceptions.
\newblock {\em Computers \& Education}, 55:1080--1090, 2010.

\bibitem{Kebritchi2017}
M.~Kebritchi, A.~Lipschuetz, and L.~Santiague.
\newblock Issues and challenges for teaching successful online courses in
  higher education.
\newblock {\em Journal of Educational Technology Systems}, 46(1):4--29, 2017.

\bibitem{Romiszowski2004}
A.~Romiszowski and R.~Mason.
\newblock Computer-mediated communication.
\newblock In H.~Jonassen, editor, {\em Handbook of research for educational
  communications and technology}. Lawrence Erlbaum, New Jersey, NJ, 2004.

\bibitem{Wise2013}
A.~F. Wise, J.~Speer, F.~Marbouti, and Y.~Hsiao.
\newblock Broadening the notion of participation in online discussions:
  Examining patterns in learners’ online listening behaviors.
\newblock {\em Instructional Science}, 41:323--343, 2013.

\bibitem{pulgar_2021}
J.~Pulgar, V.~Fahler, and A.~Spina.
\newblock Investigating how students collaborate to compose physics problems
  through structured tasks.
\newblock {\em Physical Review Physics Education Research}, 17:010120, 2021.

\bibitem{Pulgar_TSC}
J.~Pulgar.
\newblock Classroom creativity and students' social networks: theoretical and
  practical implications.
\newblock {\em Thinking Skills and Creativity}, In Press.

\bibitem{Brewe_Bruun16}
E.~Brewe, J.~Bruun, and I.~G. Bearden.
\newblock Using model analysis for multiple choice responses: A new method
  applied to force concept inventory data.
\newblock {\em Physical Review Physics Education Research}, 12:020131, 2016.

\bibitem{Borgatti2013}
S.~P. Borgatti, M.~G. Everett, and J.C. Johnson.
\newblock {\em Analyzing Social Networks}.
\newblock SAGE, Washington, DC, 2013.

\bibitem{carolan}
B.~V. Carolan.
\newblock {\em Social Network Analysis and Education: Theory, Methods and
  Applications}.
\newblock SAGE Publications Inc, 2014.

\bibitem{Grunspan}
D.~Z. Grunspan, B.~L. Wiggins, and S.~M. Goodreau.
\newblock Understanding classrooms through social network analysis: A primer
  for social network analysis in educational research.
\newblock {\em CBE-Life Sciences Education}, 13:167--178, 2014.

\bibitem{Putnik}
G.~Putnik, E.~Costa, C.~Alves, H.~Castro, L.~Varela, and V.~Shah.
\newblock Analyzing the correlation between social network analysis measures
  and performance of students in social network-based engineering education.
\newblock {\em International Journal of Technology and Design Education},
  26:413--437, 2016.

\bibitem{Bruun2013}
J.~Bruun and E.~Brewer.
\newblock Talking and learning physics: predicting future grades from network
  measures and force concept inventory pretests scores.
\newblock {\em Physical Review Physics Education Research}, 9:021109, 2013.

\bibitem{Pulgar19}
J.~Pulgar, C.~Candia, and P.~Leonardi.
\newblock Social networks and academic performance in physics: Undergraduate
  cooperation enhances ill-structured problem elaboration and inhibits
  well-structured problem solving.
\newblock {\em Physical Review Physics Education Research}, 16:010137, 2020.

\bibitem{Brewe_Sawtelle2012}
E.~Brewe, L.~Kramer, and V.~Sawtelle.
\newblock Investigating student communities with network analysis on
  interactions in a physics learning center.
\newblock {\em Physical Review Physics Education Research}, 8:010101, 2012.

\bibitem{Morris2005}
K.~V. Morris, C.~Finnegan, and W.~Sz-Shyan.
\newblock Tracking student behavior, persistence, and achievement in online
  courses.
\newblock {\em Internet and Higher Education}, 8:221--231, 2005.

\bibitem{Dawson2008}
S.~Dawson.
\newblock A study of the relationship between student social networks and sense
  of community.
\newblock {\em Educational Technology \& Society}, 11(3):224--238, 2008.

\bibitem{vyg}
L.~S. Vygotsky.
\newblock {\em Mind in Society: the Development of Higher Psychological
  Processes}.
\newblock Harvard University Press, Cambridge, MA, 1978.

\bibitem{Williams}
E.~A. Williams, J.~P. Zwolak, R.~Dou, and E.~Brewe.
\newblock Linking engagement and performance: The social network analysis
  perspective.
\newblock {\em Physical Review Physics Education Research}, 15, 2019.

\bibitem{Zwolak_Zwolak}
J.P. Zwolak, M.~Zwolak, and E.~Brewe.
\newblock Educational commitment and social networking: The power of informal
  networks.
\newblock {\em Physical Review Physics Education Research}, 14:010131, 2018.

\bibitem{Zwolak}
J.~P. Zwolak, R.~Dou, E.~A. Williams, and E.~Brewe.
\newblock Students’ network integration as a predictor of persistence in
  introductory physics courses.
\newblock {\em Physical Review Physics Education Research}, 13:010113, 2017.

\bibitem{Dou2016}
R.~Dou, E.~Brewe, J.~P. Zwolak, G.~Potvin, E.~A. Williams, and L.~H. Kramer.
\newblock Beyond performance metrics: examining a decrease in students’
  physics self-efficacy through a social network lens.
\newblock {\em Physical Review Physics Education Research}, 12:020124, 2016.

\bibitem{Carrasco2018}
C.~Carrasco, R.~Alarcón, and M.~V. Trianes.
\newblock Adatación y trabajo cooperativo en el alumnado de educación
  primaria desde la percepción del profesorado y la familia.
\newblock {\em Revista de Psicodidáctica}, 23(1):56--62, 2018.

\bibitem{leon2017}
B.~León, S.~Mendo-Lázaro, E.~Felipe-Castaño, M.I. Polo, and
  F.~Fajardo-Bullón.
\newblock Potencia de equipo y aprendizaje cooperativo en el ámbito
  universitario.
\newblock {\em Revista de Psicodidáctica}, 22(1):9--15, 2021.

\bibitem{Burt2004}
R.~S. Burt.
\newblock Structural holes and good ideas.
\newblock {\em American Journal of Sociology}, 110(2):349--399, 2004.

\bibitem{Candia}
C.~Candia, V.~Landaeta-Torres, C.~A. Hidalgo, and C.~Rodriguez-Sickert.
\newblock Strategic reciprocity improves academic performance in public
  elementary school children.
\newblock {\em Preprint arXiv:1909.11713}, 2019.

\bibitem{Granovetter}
M.S. Granovetter.
\newblock The strength of weak ties.
\newblock {\em American Journal of Sociology}, 78(6):1360--1380, 1973.

\bibitem{hansen}
M.~T. Hansen.
\newblock The search-transfer problem: The role of weak ties in sharing
  knowledge across organization subunits.
\newblock {\em Administrative Science Quarterly}, 44(1):82--111, 1999.

\bibitem{Hrastinski2009}
S.~Hrastinski.
\newblock A theory of online learning as online participation.
\newblock {\em Computers \& Education}, 52:78--82, 2009.

\bibitem{deweber}
B.~De~Wever, T.~Schellens, M.~ValckeH, and H.~Van~Keer.
\newblock Content analysis schemes to analyze transcripts of online
  asynchronous discussion groups: A review.
\newblock {\em Computers \& Education}, 46:6--28, 2006.

\bibitem{dawson2010}
S.~Dawson.
\newblock Seeing the learning community: An exploration of the development of a
  resource for monitoring online student networking.
\newblock {\em British Journal of Educational Technology}, 41, 2010.

\bibitem{traxler2020}
A.~T. Traxler, T.~Suda, E.~Brewe, and K.~Commeford.
\newblock Network positions in active learning environments in physics.
\newblock {\em Physical Review Physics Education Research}, 16:020129, 2020.

\bibitem{Commeford2020}
K.~Commeford, E.~Brewe, and A.~T. Traxler.
\newblock Characterizing active learning environments in physics using network
  analysis and copus observations.
\newblock {\em Preprint arXiv:2008.05325}, 2020.

\bibitem{Pulgar_Sochifi}
J.~Pulgar, C.~R\'{i}os, and Cristian Candia.
\newblock Physics problems and instructional strategies for developing social
  networks in university classrooms.
\newblock {\em Preprint arXiv:1904.02840}, 2019.

\bibitem{Brewe_Kramer2012}
E.~Brewe, L.~H. Kramer, and G.~E. O’Brien.
\newblock Changing participation through the formation of student learning
  communities.
\newblock In {\em AIP Conference Proceedings}, volume 1289, pages 85--88, 2012.

\bibitem{chametzky}
B.~Chametzky.
\newblock Andragogy and engagement in online learning: Tenets and solutions.
\newblock {\em Creative Education}, 5:813--821, 2014.

\bibitem{Luyt2013}
I.~Luyt.
\newblock Bridging spaces: Cross-cultural perspectives on promoting positive
  online learning experiences.
\newblock {\em Journal of Educational Technology Systems}, 42:3--20, 2013.

\bibitem{Niess2013}
M.~Niess and H.~Gillow-Wiles.
\newblock Developing asynchronous online courses: Key instructional strategies
  in a social metacognitive constructivist learning trajectory.
\newblock {\em Journal of Distance Education}, 27, 2013.

\bibitem{gupta2013}
S.~Gupta and R.~Bostrom.
\newblock Research note-an investigation of the appropriation of
  technology-mediated training methods incorporating enactive and collaborative
  learning.
\newblock {\em Information Systems Research}, 24(2):454--469, 2013.

\bibitem{merono2021}
L.~Meroño, A.~Calderón, and J.~Arias-Estero.
\newblock Pedagogía digital y aprendizaje cooperativo: efecto sobre los
  conocimientos tecnológicos y pedagógicos del contenido y el rendimiento
  académico en formación inicial docente.
\newblock {\em Revista de Psicodidáctica}, 26(1):53--61, 2021.

\bibitem{Johnson_Johnson}
D.~W. Johnson, R.~T. Johnson, and E.~J. Holubec.
\newblock {\em Circles of Learning: Cooperation in the Classroom}.
\newblock Interaction, Edina: MN, 1986.

\bibitem{Fortus}
D.~Fortus.
\newblock The importance of learning to make assumptions.
\newblock {\em Science Education}, 93(1):86--108, 2008.

\bibitem{Steiner1966}
I.~D. Steiner.
\newblock Models for inferring relationships between group size and potential
  productivity.
\newblock {\em Behavioral Science}, 11:273--283, 1966.

\bibitem{teo}
R.~Teodorescu, C.~Bennhold, G.~Feldman, and L.~Medsker.
\newblock New approach to analyzing physics problems: Ataxonomy of introductory
  physics problems.
\newblock {\em Physical Review Physics Education Research}, 9:010103, 2013.

\bibitem{Anderson2001}
L.~W. Anderson and D.R. Krathwohl.
\newblock {\em A taxonomy for learning, teaching, and assessing: A revision of
  Bloom's Taxonomy of Educational Objectives}.
\newblock Addison-Wesley/Longman, New York, USA, 2001.

\bibitem{Tinto}
V.~Tinto.
\newblock Classrooms as communities: Exploring the educational character of
  student persistence.
\newblock {\em Journal of Higher Education}, 68(6):599--623, 1997.

\bibitem{zhao2010}
L.~Zhao, Y.~Lu, B.~Wang, P.~Y. Chau, and L.~Zhang.
\newblock Cultivating the sense of belonging and motivating user participation
  in virtual communities: A social capital perspective.
\newblock {\em International Journal of Information Management},
  32(6):574--588, 2010.

\end{thebibliography}





\end{document}